\documentclass[12pt,a4paper,dvips]{article}
\usepackage{here}
\usepackage{a4p}
\usepackage{cite,mcite}
\usepackage{graphicx}

\newlength{\figwidth}
\newcommand{\etal}{{\it et~al.}}

\newcommand\GeV{\ifmmode {\mathrm{\ Ge\kern -0.1em V}}\else
                   \textrm{Ge\kern -0.1em V}\fi}%
\begin{document}
\setlength{\unitlength}{1mm}
\thispagestyle{empty}

\begin{center}
{\Large EUROPEAN ORGANIZATION FOR NUCLEAR RESEARCH}
\end{center}
\begin{flushright}
  {\large CERN-EP/98-133 }\\
  {\large August, 21 1998}\\
\end{flushright}  
\vspace{2cm}
\begin{center}
{\Large\bf Indirect Measurement of the Vertex and Angles of the Unitarity Triangle}
\vspace{2cm}

{{\bf\large  Salvatore Mele}\footnote{E-mail: Salvatore.Mele@cern.ch}\\
CERN, CH1211, Gen\`eve 23, Switzerland}
\end{center}
\vspace{2cm}
\begin{abstract}

The precise measurements of the $B^0_d$ oscillation frequency and the  limit
on the $B^0_s$ one as well as the determination of the Cabibbo-Kobayashi-Maskawa 
matrix element $|\mathrm{V_{ub}}|$ improve the constraints on the other
elements of this matrix.

A fit to the experimental data and the theory calculations leads to
the determination of the vertex of the unitarity triangle as:
\begin{displaymath}
    \rho =0.160 ^{+0.094} _{-0.070}\,\,\,\,\,
    \eta =0.381 ^{+0.061} _{-0.058}. 
\end{displaymath}
The values of its angles,  in their customary definition in terms
of sines for $\alpha$ and $\beta$, are found to be:
\begin{displaymath}
    \sin{2\alpha}  =0.06 ^{+0.35}  _{-0.42} \,\,\,\,\,
    \sin{2\beta}   =0.75 \pm 0.09 \,\,\,\,\,
    \gamma         =67 ^{+11}  _{-12}\,^\circ.
\end{displaymath}

Indirect information on  non-perturbative QCD
parameters, on the presence of a CP violating complex phase in the CKM matrix and
on the $B^0_s$ oscillation frequency are also extracted.
\end{abstract}
\begin{center}
  {\it To be submitted to Physics Letters B}
\end{center}
\newpage

%
%

\section*{Introduction}

The Standard Model~\cite{sm_glashow} of the electroweak
interactions predicts
a mixing of the quark mass eigenstates with the weak interaction ones.
This mixing is described by the Cabibbo-Kobayashi-Maskawa\cite{c}  (CKM) matrix.
Four real parameters describe this $3\times3$ unitary matrix~\cite{wolfenstein}:
\begin{equation}
  \mathrm{
    V_{CKM} = \pmatrix{ \mathrm{V_{ud}} &  \mathrm{V_{us}} &  \mathrm{V_{ub}} \cr 
      \mathrm{V_{cd}} &  \mathrm{V_{cs}} &  \mathrm{V_{cb}} \cr 
      \mathrm{V_{td}} &  \mathrm{V_{ts}} &  \mathrm{V_{tb}} \cr}
    }
  \simeq \pmatrix
  {
    1-{\lambda^2 \over 2} & \lambda &  A\lambda^3 (\rho - i\eta) \cr
    -\lambda & 1-{\lambda^2 \over 2} &  A\lambda^2 \cr
    A\lambda^3 (1-\rho - i\eta) &  -A\lambda^2 &      1 \cr
  } + {\cal{O}}(\lambda^4) .
\label{equation:ckm}
\end{equation}   
As $A$, $\rho$ and $\eta$ are of  order  unity, and
$\lambda$ is chosen as the sine of the Cabibbo angle, this
parametrisation shows immediately the hierarchy of the couplings
of the quarks in the charged current part of the  Standard Model Lagrangian.
Moreover in this parametrisation the  parameter
$\eta$ is the complex phase of the matrix and is 
thus directly related to the known violation of the CP symmetry
produced by the weak interactions.
The measurement of the parameters of the CKM matrix is thus of 
fundamental importance for both the precision description of the
weak interaction of quarks and the investigation of the mechanism of
CP violation.

The parameters $A$ and $\lambda$ are known with an accuracy of a few
percent and the determination
of $\rho$ and $\eta$ is the subject of this letter. A large number
of physical processes can be parametrised in terms of the values of
the elements of the CKM matrix, together with other parameters of theoretical
and experimental origin. Four of them  show good sensitivity for the indirect
determination of $\rho$ and $\eta$ and
are discussed in what follows.
A fit based on this information has been performed, as suggested in~\cite{ali1},
and its results are  presented below.

As it is well known the measurement of  $\rho$ and $\eta$ 
is equivalent to the determination of the only unknown vertex and the angles
of a triangle in the $\rho-\eta$ plane whose other two vertices are in (0,0) and (1,0). 
Figure~1 shows this triangle, called the unitarity triangle.

%
%

\section*{Constraints}

The value of the sine of the Cabibbo angle  is known with a good accuracy~\cite{pdg} as: 
\begin{displaymath}
\lambda = 0.2196 \pm 0.0023.
\end{displaymath}
The  parameter $A$ depends on $\lambda$ and on the CKM matrix element $|\mathrm{V_{cb}}|$.
Using the value~\cite{pdg}:
\begin{displaymath}
 |\mathrm{V_{cb}}| = (39.5 \pm 1.7)\times 10^{-3},
\end{displaymath}
it can be extracted:
\begin{displaymath}
A = {|\mathrm{V_{cb}}|^2 \over \lambda^2} = 0.819 \pm 0.035.
\end{displaymath}

The four processes most sensitive to the value of the CKM parameters
$\rho$ and $\eta$ are described in the following, along with their experimental 
knowledge and  theoretical dependences. 

%
%

\subsection*{CP Violation for Neutral Kaons}

The mass eigenstates of the
neutral kaons can be written as
$|K_S\rangle = p |K^0\rangle + q |\bar{K^0}\rangle$ and 
$|K_L\rangle = p |K^0\rangle - q |\bar{K^0}\rangle$.
The relation $p\neq q$ implies the violation of CP that, in the 
Wu-Yang phase convention~\cite{wuyang}, is described by the parameter $\epsilon_K$
defined as:
\begin{displaymath}
{p \over q} = {1 + \epsilon_K \over 1 - \epsilon_K  }.
\end{displaymath}
The precise measurements of the $K_S \rightarrow \pi^+ \pi^-$ and 
$K_L \rightarrow \pi^+ \pi^-$ decay rates imply~\cite{pdg}:
\begin{displaymath}
|\epsilon_K| = (2.280 \pm 0.019) \times 10^{-3}.
\end{displaymath}
The relation of $|\epsilon_K|$ to the CKM matrix parameters is~\cite{buras1,buras2}:
\begin{equation}
|\epsilon_K| = {G_F^2 f^2_K m_K m_W^2 \over 6 \sqrt{2} \pi^2 \Delta m_K}
 B_K \left( A^2 \lambda^6 \eta \right) \big[ y_c \left( \eta_{ct}f_3(y_c,y_t)-  \eta_{cc}\right) 
 + \eta_{tt} y_t f_2(y_t) A^2 \lambda^4 \left(1-\rho\right)\big].                          
\label{equation:ek}
\end{equation}
The functions $f_3$ and $f_2$ of the variables
$y_t = m_t^2 / m_W^2$ and $y_c = m_c^2 / m_W^2$ are given by~\cite{ali1}:
\begin{equation}
f_2(x) = {1 \over 4} + {9 \over 4(1-x)} - {3 \over (1-x)^2} - {3 x^2 \ln{x} \over 2(1-x)^3},\,\,\,\,
f_3(x,y) = \ln{y \over x} -{3y \over 4(1-y)}\left( 1 + {y \ln{y}\over 1-y}\right).
\label{equation:f2}
\end{equation}

From the value of the mass of the top quark reported by the CDF and D0 
collaborations~\cite{pdg}, $173.8 \pm 5.2\,\GeV$, and the scaling
proposed in~\cite{buras3} one obtains:

\begin{displaymath}
\overline{m_t}(m_t) = 166.8 \pm 5.3\,\GeV,
\end{displaymath}
while the mass of the charm quark is~\cite{pdg}:
\begin{displaymath}
\overline{m_c}(m_c) = 1.25 \pm 0.15\,\GeV.
\end{displaymath}
The current values of the parameters that include the calculated QCD corrections 
are~\cite{buras3,buras4}:
\begin{displaymath}
\eta_{cc} = 1.38 \pm 0.53,\,\,\,\,
\eta_{tt} = 0.574 \pm 0.004,\,\,\,\,
\eta_{ct} = 0.47 \pm 0.04.
\end{displaymath}
The largest theoretical uncertainty which affects this constraint is that on the ``bag'' 
parameter $B_K$, that reflects non-perturbative QCD contributions to the process. 
Using the value of the JLQCD collaboration~\cite{JLQCD},
$B_K(2\,\GeV) = 0.628\pm 0.042$, with a calculation similar to that
reported in~\cite{sharpe1} the  value  used in the following can be derived as:
\begin{displaymath}
B_K  = 0.87\pm 0.14.
\end{displaymath}

The other physical constants of the formula are reported in Table~1. This constraint
has the shape of an hyperbola in the $\rho-\eta$ plane.

%
%

\subsection*{Oscillations of $\boldmath{B^0_d}$ Mesons}

Neutral mesons containing a $b$ quark show a behaviour similar to neutral kaons. 
The heavy and light mass eigenstates, $B_L$ and $B_H$ respectively,
are different from the CP eigenstates $B^0_d$ and $\bar{B^0_d}$:
\begin{displaymath}
|B_L\rangle = p |B^0_d\rangle + q |\bar{B^0_d}\rangle 
\,\,\,\,\,\,\,\,\,\,
|B_H\rangle = p |B^0_d\rangle - q |\bar{B^0_d}\rangle.
\end{displaymath}
In the neutral B system the mass difference $\Delta m_d = m_{B_H} -  m_{B_L}$ 
is the key feature of the physics  while the lifetime difference dominates the 
effects in the neutral kaon system.
This mass difference can be measured by means of the study of the oscillations
of one CP eigenstate into the other. The high precision world average is~\cite{alexander}:
\begin{displaymath}
\Delta m_d            = 0.471 \pm 0.016\,\mathrm{ps}^{-1}.
\end{displaymath}
The relation of $\Delta m_d$ with the CKM parameters, making
use of the Standard Model description of the box diagrams that
give rise to the mixing and the parametrisation~(\ref{equation:ckm}) of the CKM 
matrix, reads:
\begin{equation}
\Delta m_d = {G_F^2 \over 6 \pi ^2}m_W^2 m_B \left(f_{B_d}\sqrt{B_{B_d}}\right)^2
\eta_B y_t f_2(y_t) A \lambda^6 \left[ \left(1-\rho\right)^2 + \eta^2\right].
\label{equation:dmd}
\end{equation}
The function $f_2$ is given by~(\ref{equation:f2}), the value of the
calculated QCD correction $\eta_B$
is~\cite{buras3,buras4}:
\begin{displaymath}
     \eta_B  = 0.55 \pm 0.01,
\end{displaymath}
and the equivalent of the $B_K$ parameter for the kaon system, $f_{B_d}\sqrt{B_{B_d}}$,
is taken as~\cite{flynn2}:
\begin{displaymath}
     f_{B_d}\sqrt{B_{B_d}} = 0.201 \pm 0.042\,\GeV.
\end{displaymath}

The measurement of $\Delta m_d$ constrains the vertex of the unitarity triangle to a
circle in the $\rho-\eta$ plane, centred in $(1,0)$.

\subsection*{Oscillations of $\boldmath{B^0_s}$ Mesons}

$B^0_s$ mesons are believed to undergo a mixing analogous to the 
$B^0_d$ ones. Their larger mass difference 
$\Delta m_s$ is responsible for oscillations that are faster than
the $B^0_d$ ones, and have thus still eluded direct observation.
A lower limit has been set by the LEP B oscillation working group 
combining the results of the searches performed by the
LEP experiments with a contribution from the SLD and CDF collaborations, as~\cite{parodi}:
\begin{displaymath}
  \Delta m_s > 12.4\,\mathrm{ps}^{-1}\,\,\, (95\%\,\mathrm{C.L.}).
\end{displaymath}

The expression for $\Delta m_s$ in the Standard Model is similar to that
for  $\Delta m_d$. From the ratio of these two expressions the value of 
$\Delta m_s$ can be written as:
\begin{equation}
\Delta m_s = \Delta m_d {1 \over \lambda^2}{m_{B_s} \over m_{B_d}} \xi^2 
{1 \over \left( 1- \rho\right)^2 + \eta^2},
\label{equation:dms}
\end{equation}
where all the theoretical uncertainties are included in the  quantity
$\xi$, known as~\cite{flynn2}:
\begin{displaymath}
\xi = { f_{B_d}\sqrt{B_{B_d}} \over f_{B_s}\sqrt{B_{B_s}} } = 1.14 \pm 0.08.
\end{displaymath}

This experimental lower limit excludes the  values of the vertex
of the unitarity triangle outside  a circle in the $\rho-\eta$ plane  with centre in $(1,0)$.


\subsection*{Charmless Semileptonic b Decays}

The three constraints described above are all affected by a large
theoretical uncertainty on some of the parameters that enter their expression, 
namely $B_K$, $f_{B_d}\sqrt{B_{B_d}}$ and $\xi$. A determination of either 
$|\mathrm{V_{ub}}|$ or the ratio $|\mathrm{V_{ub}}|/|\mathrm{V_{cb}}|$ allows
a more sensitive constraint not relaying  on any  non-perturbative QCD calculation.
It follows from the CKM matrix parametrisation of~(\ref{equation:ckm}) that:
\begin{equation}
 |\mathrm{V_{ub}}|/|\mathrm{V_{cb}}| = \lambda \sqrt{\rho^2 + \eta^2}.
\label{equation:vubvcb}
\end{equation}

The CLEO collaboration has measured this ratio by means
of the endpoint of inclusive~\cite{cleoinc2}  charmless semileptonic B decays as:
$|\mathrm{V_{ub}}|/|\mathrm{V_{cb}}| = 0.08 \pm 0.02$.
The ALEPH and L3 collaborations  have recently 
measured at LEP the  inclusive charmless semileptonic branching
fraction of beauty hadrons, $\mathrm{Br}(b\rightarrow X_u\ell\nu)$, from
which the value of $|\mathrm{V_{ub}}|$ can be extracted~\cite{uraltsev} as:
\begin{equation}
  |\mathrm{V_{ub}}| = 0.00458 \times \sqrt{ 
{\mathrm{Br}(b\rightarrow X_u\ell\nu)}
\over 0.002 } \times  \sqrt{ 1.6\,\mathrm{ps} \over \tau_B  } \pm 4\%_{\mathrm{theory}}.
\label{equation:vub}
\end{equation}
The experimental results are :
\begin{center}
\begin{tabular}{rl}
ALEPH~\cite{alephvub}: & Br$(b\rightarrow X_u\ell\nu) = (1.73 \pm 0.55 \pm 0.55) \times 10^{-3}$\\
L3~\cite{l3vub}:    & Br$(b\rightarrow X_u\ell\nu) = (3.3 \pm 1.0 \pm 1.7) \times 10^{-3}$,\\
\end{tabular}
\end{center}
where the first uncertainty is statistical and the second systematic, with the average:
\begin{displaymath}
 \mathrm{Br}(b\rightarrow X_u\ell\nu) = (1.85 \pm 0.52 \pm 0.59 )\times 10^{-3},
\end{displaymath}
with the same meaning of the uncertainties. This value makes it possible to determine
$|\mathrm{V_{ub}}|$  at LEP by means of the formula~(\ref{equation:vub}) as:
\begin{displaymath}
|\mathrm{V_{ub}}| = (4.5  \,^{+0.6}_{-0.7}\, ^{+0.7}_{-0.8} \pm 0.2)\times 10^{-3}.
\end{displaymath}
The first uncertainty is statistical, the second systematic and the third 
theoretical. The value $\tau_B = (1.554 \pm 0.013)\mathrm{ps}$~\cite{bately} has been used.
Using the quoted value of $|\mathrm{V_{cb}}|$ the combination with the CLEO measurement
yields:
\begin{displaymath}
|\mathrm{V_{ub}}|/|\mathrm{V_{cb}}| = 0.093 \pm 0.016.
\end{displaymath}
The uncertainty on this important constraint is thus significantly reduced by the inclusion
of the recent LEP measurements. A further reduction to 0.015 could be achieved by the inclusion
of the DELPHI collaboration preliminary  measurement of this quantity~\cite{delphivancouver}.

This constraint gives a  circle in the $\rho-\eta$ plane with centre in (0,0),
shown in Figure~2 together with  all the other constraints described above.

%
%
\section*{Determination of $\boldmath{\rho}$ and $\boldmath{\eta}$}

The $\rho$ and $\eta$ parameters can be determined from a fit to
the experimental values of all the constraints described above.
The experimental and theoretical quantities that appear in the formulae 
describing the constraints have
been fixed to their central values if their errors were reasonably small, and are
reported in the left half of~Table~1. The quantities affected by a larger
error have been used as additional parameters of the fit, but including a constraint
on their value. 
This procedure has been implemented making use of the MINUIT package~\cite{minuit}
to minimise the following expression:

\begin{displaymath}
\chi^2 =  {\left(\widehat{A} - A\right)^2 \over \sigma_{A}^2} +   
          {\left(\widehat{m_c} - m_c\right)^2 \over \sigma_{m_c}^2} +
          {\left(\widehat{m_t} - m_t\right)^2 \over \sigma_{m_t}^2} + 
          {\left(\widehat{B_K} - B_K\right)^2 \over \sigma_{B_K}^2} +
\end{displaymath}
\begin{displaymath}
          +{\left(\widehat{\eta_{cc}} -\eta_{cc} \right)^2 \over \sigma_{\eta_{cc}}^2} + 
           {\left(\widehat{\eta_{ct}} -\eta_{ct} \right)^2 \over \sigma_{\eta_{ct}}^2} + 
           {\left(\widehat{f_{B_d}\sqrt{B_{B_d}}} - f_{B_d}\sqrt{B_{B_d}}\right)^2 
                 \over \sigma_{f_{B_d}\sqrt{B_{B_d}}}^2} +
           {\left(\widehat{\xi} - \xi\right)^2 \over \sigma_{\xi}^2} + 
           {\left(\widehat{|\mathrm{V_{ub}}|\over|\mathrm{V_{cb}}|} 
               - {|\mathrm{V_{ub}}|\over|\mathrm{V_{cb}}|}\right)^2 \over \
             \sigma_{{|\mathrm{V_{ub}}| \over |\mathrm{V_{cb}}|}}^2} +
\end{displaymath}
\begin{displaymath}
          +{\left(\widehat{|\epsilon_K|} - |\epsilon_K| \right)^2  
          \over  \sigma_{|\epsilon_K|}^2} +
           {\left(\widehat{\Delta m_d} - \Delta m_d\right)^2
          \over  \sigma_{\Delta m_d}^2}+
          \chi^2\left( {\cal{A}}\left(\Delta m_s\right), \sigma_{\cal{A}}\left(\Delta m_s \right)\right).
\end{displaymath}

The symbols with a hat represent the reference values measured or 
calculated for a given physical quantity, as listed
in Table~1, while the corresponding $\sigma$ are their errors. 
The parameters of the fit are $\rho$, $\eta$, $A$, $m_c$, $m_t$, $B_K$, $\eta_{ct}$, 
$\eta_{cc}$, $f_{B_d}\sqrt{B_{B_d}}$ and $\xi$, that are used to calculate
the values of $|\epsilon_K|$,  $\Delta m_d$,  $\Delta m_s$ and $|\mathrm{V_{ub}}|\over|\mathrm{V_{cb}}|$
by means of the formulae~(\ref{equation:ek}),~(\ref{equation:dmd}),~(\ref{equation:dms}) 
and~(\ref{equation:vubvcb}).

As no measurements of $\Delta m_s$ are available a further contribution to the
$\chi^2$ analogous to the previous ones can not be calculated. The following approximation
has been used to extract a contribution 
from the Confidence Levels of the $\Delta m_s$ exclusion.
The results of the search for $B^0_s$ oscillations have
been presented and combined~\cite{alexander} in terms of the oscillation amplitude ${\cal{A}}$~\cite{moser},
a parameter that is zero in the absence of signal and compatible with one if an
oscillation signal is observed, as in:
\begin{displaymath}
P \left[ B^0_s \rightarrow (B^0_s, \bar{B^0_s}) \right]
= {1 \over 2 \tau_s} e^{-t/\tau_s} \left( 1 \pm {\cal {A}} \cos{\Delta m_s}
\right).
\end{displaymath}
The experimental results are reported in terms of  ${\cal{A}}\left(\Delta m_s\right)$
and $\sigma_{\cal{A}}\left(\Delta m_s \right)$, which leads to the quoted 95\% Confidence
Level limit as the value of $\Delta m_s$ for which the area above one of the Gaussian distribution
with mean ${\cal{A}}\left(\Delta m_s\right)$ and variance $\sigma^2_{\cal{A}}\left(\Delta m_s \right)$
equals the 5\% of the total area.
As noted in~\cite{paganini} the full set of combined ${\cal{A}}\left(\Delta m_s\right)$
and $\sigma_{\cal{A}}\left(\Delta m_s \right)$ measurements indeed contains more information 
than this limit and it is used in this procedure, with a different statistical approach. The value of
$\Delta m_s$ can be calculated for each value taken by the fit parameters  $\rho$, $\eta$ and $\xi$
by means of formula~(\ref{equation:dms}), together with the value of its corresponding
Confidence Level obtained as described above. The value 
$\chi^2\left( {\cal{A}}\left(\Delta m_s\right), \sigma_{\cal{A}}\left(\Delta m_s \right)\right)$
of a $\chi^2$ distribution with one degree of freedom corresponding to this Confidence Level can then
be calculated and added to the total $\chi^2$ of the fit.

\begin{table}[t] 
  \begin{center}
    \begin{tabular}{|rcl|rcl|}
     \hline
     &  &  Fixed in the fit &                                                & & Varied in the fit \\                                          
     \hline                                                                                                                                   
     $\lambda               $& = & $0.2196 \pm 0.0023$ &                                 $A                     $& = & $0.819\pm 0.035$   \\               
     $G_F                   $& = & $(1.16639 \pm 0.00001)\times 10^{-5}$\,\GeV$^{-2}$&   $\eta_{ct}             $& = & $0.47 \pm 0.04$   \\                
     $f_K                   $& = & $0.1598 \pm 0.0015$\,\GeV &                           $\eta_{cc}             $& = & $1.38 \pm 0.53$  \\                 
     $\Delta m_K            $& = & $(0.5304 \pm 0.0014)\times 10^{-2}$\,ps$^{-1}$ &      $\overline{m_c}(m_c)   $& = & $1.25 \pm 0.15$\,\GeV   \\          
     $m_K                   $& = & $0.497672 \pm 0.000031$\,\GeV &                       $\overline{m_t}(m_t)   $& = & $166.8 \pm 5.3$\,\GeV  \\           
     $m_W                   $& = & $80.375 \pm 0.064$\,\GeV &                            $f_{B_d}\sqrt{B_{B_d}} $&=& $0.201 \pm 0.042$\,\GeV  \\           
     $m_{B_d}               $& = & $5.2792 \pm 0.0018$\,\GeV &                           $B_K                   $&=& $0.87 \pm 0.14$ \\                    
     $m_{B_s}               $& = & $5.3692 \pm 0.0020$\,\GeV &                           $\xi                   $&=& $1.14 \pm 0.08$ \\                    
     $m_B                   $& = & $5.290 \pm 0.002$\,\GeV &                             $|\epsilon_K|          $& = & $(2.280\pm 0.019)\times 10^{-3}$  \\
     $\eta_B                $& = & $0.55 \pm 0.01$ &                                     $\Delta m_d            $& = & $0.471 \pm 0.016$\,ps$^{-1}$ \\     
     $\eta_{tt}             $& = & $0.574 \pm 0.004 $&                                   $|\mathrm{V_{ub}}|/|\mathrm{V_{cb}}| $& = & $0.093 \pm 0.016 $ \\ 
     \hline    
   \end{tabular}
   \caption{Physical constants and parameters of the fit. The values whose
     origin is not discussed in the text are from~[5].}
 \end{center}
\end{table}

The results of the fit are the following:
\begin{displaymath}
    \rho =0.160 ^{+0.094} _{-0.070}\,\,\,\,\,
    \eta =0.381 ^{+0.061} _{-0.058}. 
\end{displaymath}
The 95\% Confidence Level regions for $\rho$ and $\eta$ are:
\begin{displaymath}
    -0.02 < \rho < 0.35 \,\,\,\,\,
     0.27 < \eta < 0.50 \,\,\,(\mathrm{95\% C.L.}).
\end{displaymath}
Figure~3a shows the allowed confidence regions in the $\rho-\eta$ plane, together
with the favoured unitarity triangle, that is also shown superimposed on the constraints
of Figure~2.

From these results it is possible to determine also the value of the angles
of the unitarity triangle. The angles $\alpha$ and $\beta$ are reported in terms of the functions
$\sin{2\alpha}$ and $\sin{2\beta}$ as will be measured at the next B-factories. 
The numerical values obtained from the fit are:
\begin{displaymath}
    \sin{2\alpha}  =0.06 ^{+0.35}  _{-0.42} \,\,\,\,\,
    \sin{2\beta}   =0.75 \pm 0.09 \,\,\,\,\,
    \gamma         =67 ^{+11}  _{-12}\,^\circ.
\end{displaymath}
In terms of 95\% Confidence Level regions these last results can be
expressed as:
\begin{displaymath}
   -0.71 <  \sin{2\alpha} < 0.70 \,\,\,\,\,
   0.56 <  \sin{2\beta}  < 0.94 \,\,\,\,\,
   44^\circ  <   \gamma       < 93^\circ  \,\,\,(\mathrm{95\% C.L.})  
\end{displaymath}

%
%
\section*{Interpretations}

The fit procedure described above can also  be used  to extract information on the
theory parameters that enter the fit with a large uncertainty and at the same
time, perform an estimation of $\rho$ and $\eta$ independent of them. This can be achieved
by removing  from the fit the constraint on the parameter.
The two parameters $B_K$ and $f_{B_d}\sqrt{B_{B_d}}$ are those affected by the 
largest theory uncertainty. By applying this method to the parameter $B_K$, the
fit yields:
\begin{displaymath}
    \rho =0.156 ^{+0.096} _{-0.091} \,\,\,\,\,
    \eta = 0.393 ^{+0.072} _{-0.080} \,\,\,\,\, 
    B_K  = 0.80 ^{+0.27} _{-0.16}.
\end{displaymath}
The value of $B_K$ favoured by the fit has an error larger than that on the
estimated input parameter and thus can not help in restricting its range
of allowed values.
The same procedure with  $f_{B_d}\sqrt{B_{B_d}}$ as a free parameter leads to 
the results:
\begin{displaymath}
    \rho =0.186 ^{+0.085} _{-0.093}  \,\,\,\,\,
    \eta = 0.379 ^{+0.061} _{-0.057} \,\,\,\,\,
    f_{B_d}\sqrt{B_{B_d}} = 0.222 ^{+0.026} _{-0.011}\,\GeV, 
\end{displaymath}
the value of $f_{B_d}\sqrt{B_{B_d}}$ comes out to be well in agreement
with the predicted one with a smaller uncertainty.
The same procedure applied to $B_K$ and $f_{B_d}\sqrt{B_{B_d}}$ simultaneously
gives:
\begin{displaymath}
    \rho =0.170 ^{+0.320} _{-0.098}\,\,\,\,\,
    \eta = 0.390 ^{-0.059} _{-0.103}\,\,\,\,\, 
    B_K  = 0.82 ^{+0.41} _{-0.17} \,\,\,\,\, 
    f_{B_d}\sqrt{B_{B_d}} = 0.217 ^{+0.047} _{-0.022}\,\GeV.
\end{displaymath}

The $\Delta m_s$ constraint has a big impact on the  $\rho$ 
uncertainty as can be observed by removing it from the fit, what gives:
\begin{displaymath}
    \rho =0.022 _{-0.264} ^{+0.145}\,\,\,\,\,
    \eta =0.434 _{-0.090} ^{+0.062}. 
\end{displaymath}
Figure~3b shows the experimentally favoured regions in the $\rho-\eta$ 
plane for this fit together with the lower limit and expected sensitivity 
($\Delta m_s = 13.8\,\mathrm{ps}^{-1}$~\cite{parodi})
of the current experiments to $B^0_s$ oscillations. The confidence regions
for $\Delta m_s$ can be extracted from this fit as:
\begin{displaymath}
    \Delta m_s = 11.3 _{-3.9} ^{+3.0}\,\mathrm{ps}^{-1} \\
\end{displaymath}
\begin{displaymath}
     5.7\,\mathrm{ps}^{-1} < \Delta m_s   < 17.8\,\mathrm{ps}^{-1} \,\,\,(\mathrm{95\% C.L.}),
\end{displaymath}

The LEP measurements have greatly improved the constraints on the
CKM matrix. 
Another fit has been performed removing the $\Delta m_s$ constraint,
derived mainly from the  LEP limits, and excluding the LEP
measurement from the averages of the other input quantities; that is
using:
\begin{eqnarray*}
|\mathrm{V_{ub}}|/|\mathrm{V_{cb}}| &=& 0.08 \pm 0.02\\
|\mathrm{V_{cb}}|   &=&(50\pm 5)\times 10^{-3} \\
\Delta m_d              &=&(0.500 \pm 0.030)\,\mathrm{ps}^{-1}.
\end{eqnarray*}
The first value is that quoted above from the CLEO collaboration~\cite{cleoinc2},
the second follows from~\cite{patterson} and the last  has been 
estimated from the current published and preliminary
results from the CDF and SLD collaborations.
This fit, as shown in  Figure~3c, yields:
\begin{displaymath}
    \rho =0.012 ^{+0.192} _{-0.252}\,\,\,\,\,
    \eta =0.383 ^{+0.082} _{-0.093}\,\,\,\mathrm{and}
\end{displaymath}
\begin{displaymath}
    \sin{2\alpha}  = 0.63  ^{+0.37}  _{-0.90} \,\,\,\,\,
    \sin{2\beta}   = 0.67  ^{+0.14}  _{-0.22} \,\,\,\,\,
    \gamma         = 88 ^{+39}  _{-28}\,^\circ .
\end{displaymath}
Some of the errors are reduced by as much as a factor three
by the inclusion of the LEP data.
%
%
\section*{A real CKM matrix ?}

To date the only experimental evidence for the violation of CP in
the CKM matrix, namely its complex phase described
by a value of $\eta$ different from zero, comes from the neutral kaon system.
As different models have been proposed to explain that effect, it
is of interest to remove from the fit the constraint
related to this process and then investigate the compatibility of
$\eta$ with zero~\cite{barbieri}. This procedure yields the following results, graphically displayed in Figure~3d:
\begin{displaymath}
    \rho =0.156 _{-0.090} ^{+0.096}\,\,\,\,\,
    \eta =0.394 _{-0.080} ^{+0.072}. 
\end{displaymath}
The value of $\eta$ is not compatible with zero at the 95\% and 99\% of
Confidence Levels either:
\begin{eqnarray*}
    -0.025 < \rho < 0.358 \,\,\,(\mathrm{95\% C.L.})&\,\,\,\,\,\,      -0.069 < \rho < 0.411 \,\,\,(\mathrm{99\% C.L.})\\  
    0.224 < \eta < 0.531 \,\,\,(\mathrm{95\% C.L.})& \,\,\,\,\,\,     \phantom{-}0.157 < \eta < 0.574 \,\,\,(\mathrm{99\% C.L.}).
\end{eqnarray*}

If the CKM matrix is assumed to be real, as recently proposed for instance in~\cite{glashow},
all the circular constraints reduce to linear intervals on the $\rho$ axis, onto which
the unitarity triangle will then be projected. 
This hypothesis can be checked removing again the neutral kaon constraints from the fit and
modifying the formulae~(\ref{equation:dmd}),~(\ref{equation:dms}) 
and~(\ref{equation:vubvcb}) imposing $\eta$ equal to zero. The result
of this fit, whose parameters are reduced to $\rho$,  $A$,  $m_t$,  $f_{B_d}\sqrt{B_{B_d}}$ and $\xi$, is:
\begin{displaymath}
    \rho =0.321 _{-0.056} ^{+0.053}.
\end{displaymath}
The value of the $\chi^2$ function at the minimum is 6.7, leading to the
conclusion that a CKM matrix real by construction can  fit the data.

%
%
\section*{Conclusions}

The combination of the precise  measurements of $\Delta m_d$, the updated limits
on $\Delta m_s$ and the determination of  $|\mathrm{V_{ub}}|$ 
helps in constraining the CKM matrix elements.

From a simultaneous fit to all the available data and 
theory parameters 
the vertex of the unitarity triangle is determined as:
\begin{displaymath}
    \rho =0.160 ^{+0.094} _{-0.070}\,\,\,\,\,
    \eta =0.381 ^{+0.061} _{-0.058}. 
\end{displaymath}
yielding the following values for its angles:
\begin{displaymath}
    \sin{2\alpha}  =0.06 ^{+0.35}  _{-0.42} \,\,\,\,\,
    \sin{2\beta}   =0.75 \pm 0.09 \,\,\,\,\,
    \gamma         =67 ^{+11}  _{-12}\,^\circ.
\end{displaymath}

The accuracy on $\sin{2\beta}$ from this indirect analysis is already 
at the same level as that expected to be achieved with the direct measurement 
at the B-factories due to become operational in the next future.
These limits greatly benefit from the inclusion of LEP data. 

The fit suggests the value of
the non-perturbative QCD parameter $f_{B_d}\sqrt{B_{B_d}}$ as: 
\begin{displaymath}
    f_{B_d}\sqrt{B_{B_d}} = 0.222 ^{+0.026} _{-0.011}\,\GeV.
\end{displaymath}

The parameter $\eta$ related to the complex phase of the matrix and
thus to the CP violation is found to be different from zero at more than the
99\% Confidence Level, 
even removing  from the fit the constraints arising
from CP violation in the neutral kaon
system. Nonetheless the hypothesis of a real matrix can still fit the
data without this constraint.

The fit also indicates  the $\Delta m_s$ variation range as:
\begin{displaymath}
    \Delta m_s = (11.3 _{-3.9} ^{+3.0})\,\mathrm{ps}^{-1} 
\end{displaymath}
\begin{displaymath}
     5.7\,\mathrm{ps}^{-1} < \Delta m_s  < 17.8\,\mathrm{ps}^{-1} \,\,\,(\mathrm{95\% C.L.}).
\end{displaymath}

These results improve those of similar previous analyses
~\cite{paganini,ali2} and agree
with another one based on a different approach~\cite{parodi}.

%
\section*{Acknowledgements}

I would like to thank Joachim Mnich for the interesting discussions on the fit
procedures and John Field for his careful reading of this manuscript.
I am grateful to Sheldon Glashow for having suggested to me to fit a real CKM matrix.

%
%

\newpage
%
%

\begin{figure}[p]
  \begin{center}
      \mbox{\includegraphics[width=0.5\figwidth]{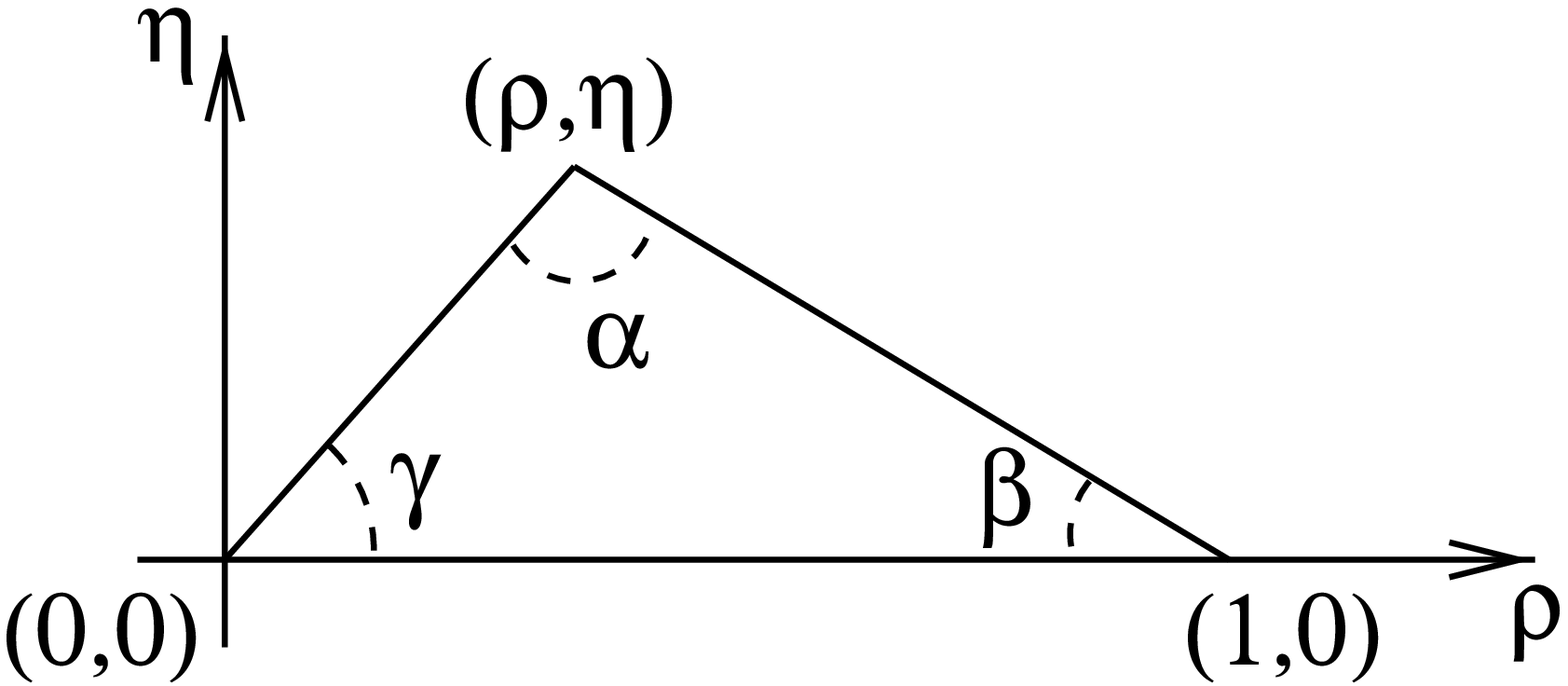}}
    \caption{The unitarity triangle.}
  \end{center}
\end{figure}

\begin{figure}[p]
  \begin{center}
      \mbox{\includegraphics[width=\figwidth]{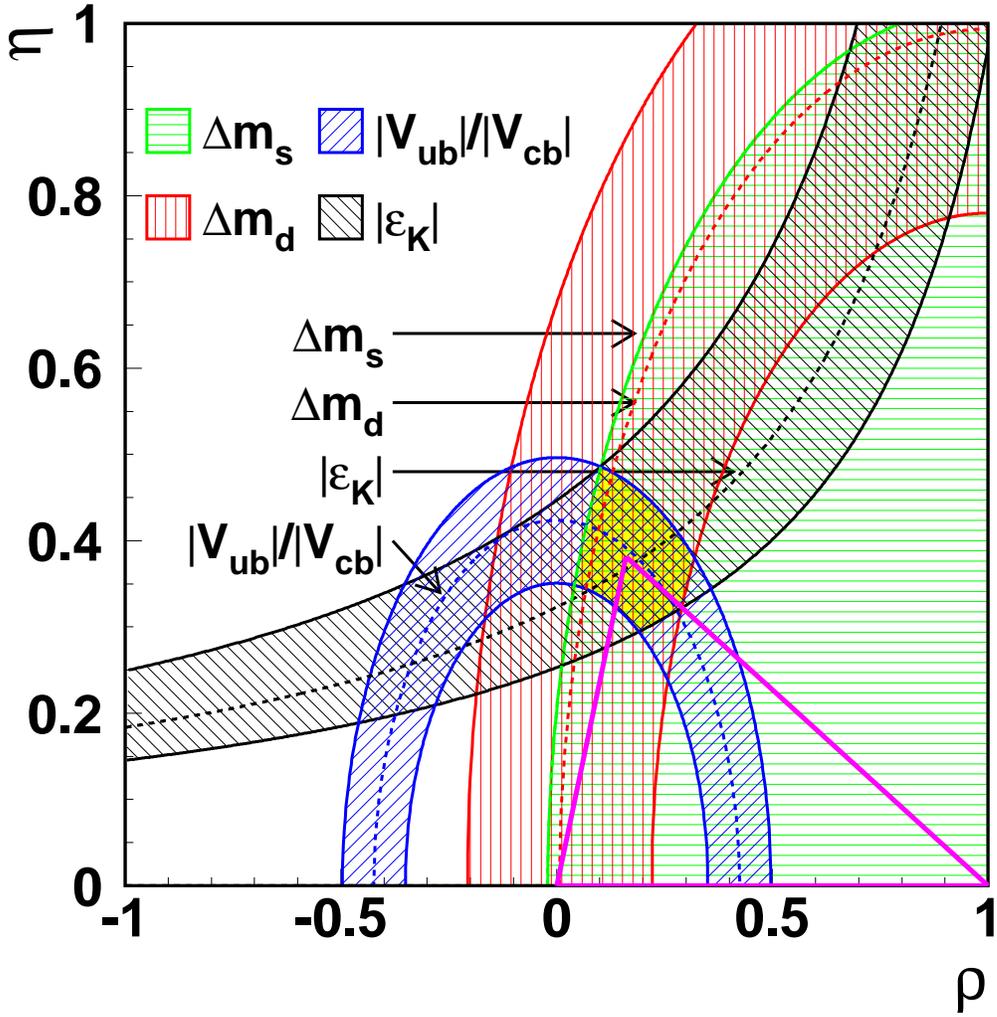}}
    \caption{The current constraints and the favoured unitarity triangle. The constraint
      coming from $\mathrm{B}^0_\mathrm{s}$ oscillations is a limit at 95\% of Confidence
      Level, while the others represent a $\pm 1 \sigma$ variation of the experimental
      and theoretical parameters entering the formulae in the text.}
  \end{center}
\end{figure}

\begin{figure}[p]
  \begin{center}
    \begin{tabular}{cc}
      \mbox{\includegraphics[width=0.5\figwidth]{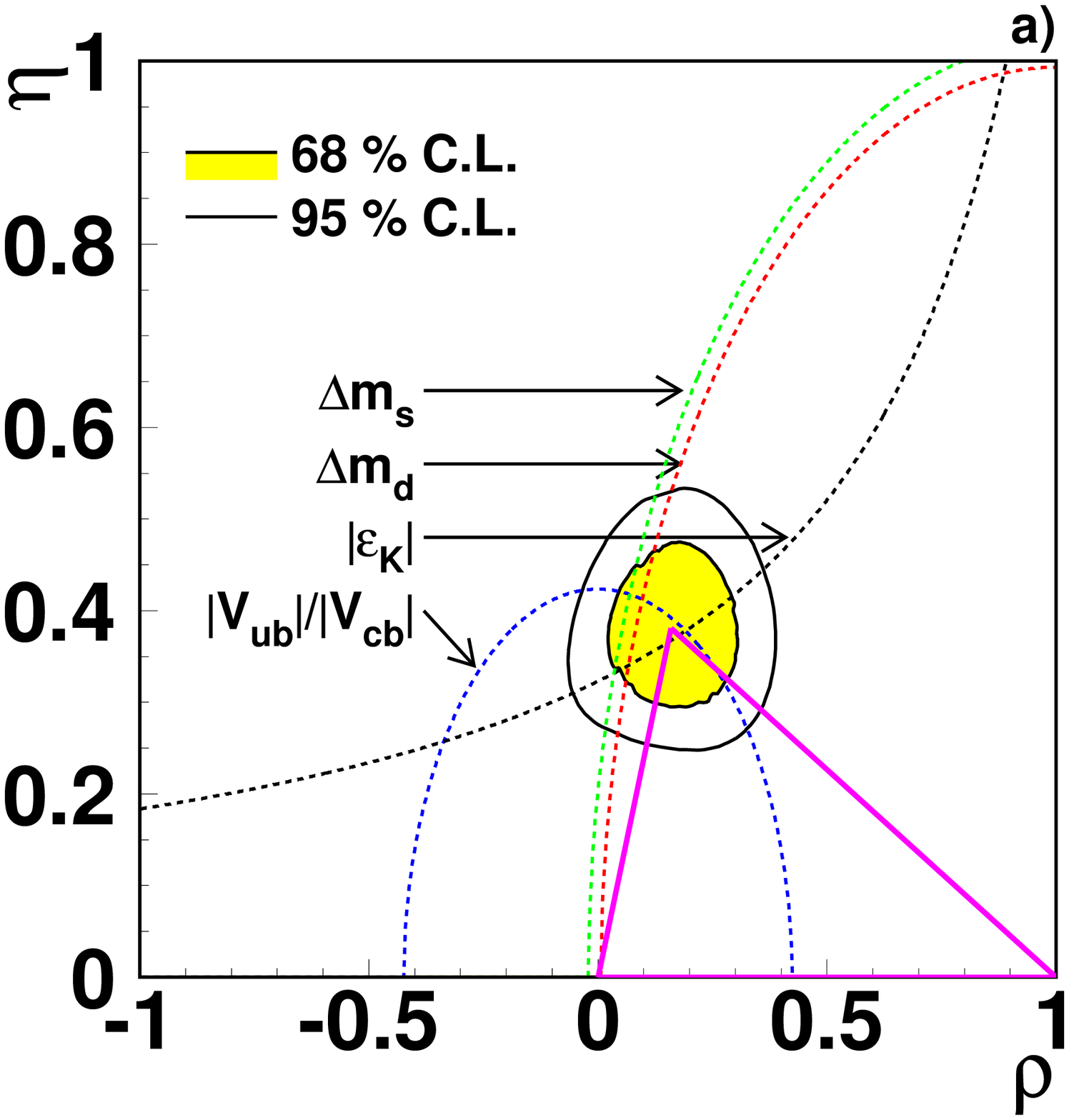}} &
      \mbox{\includegraphics[width=0.5\figwidth]{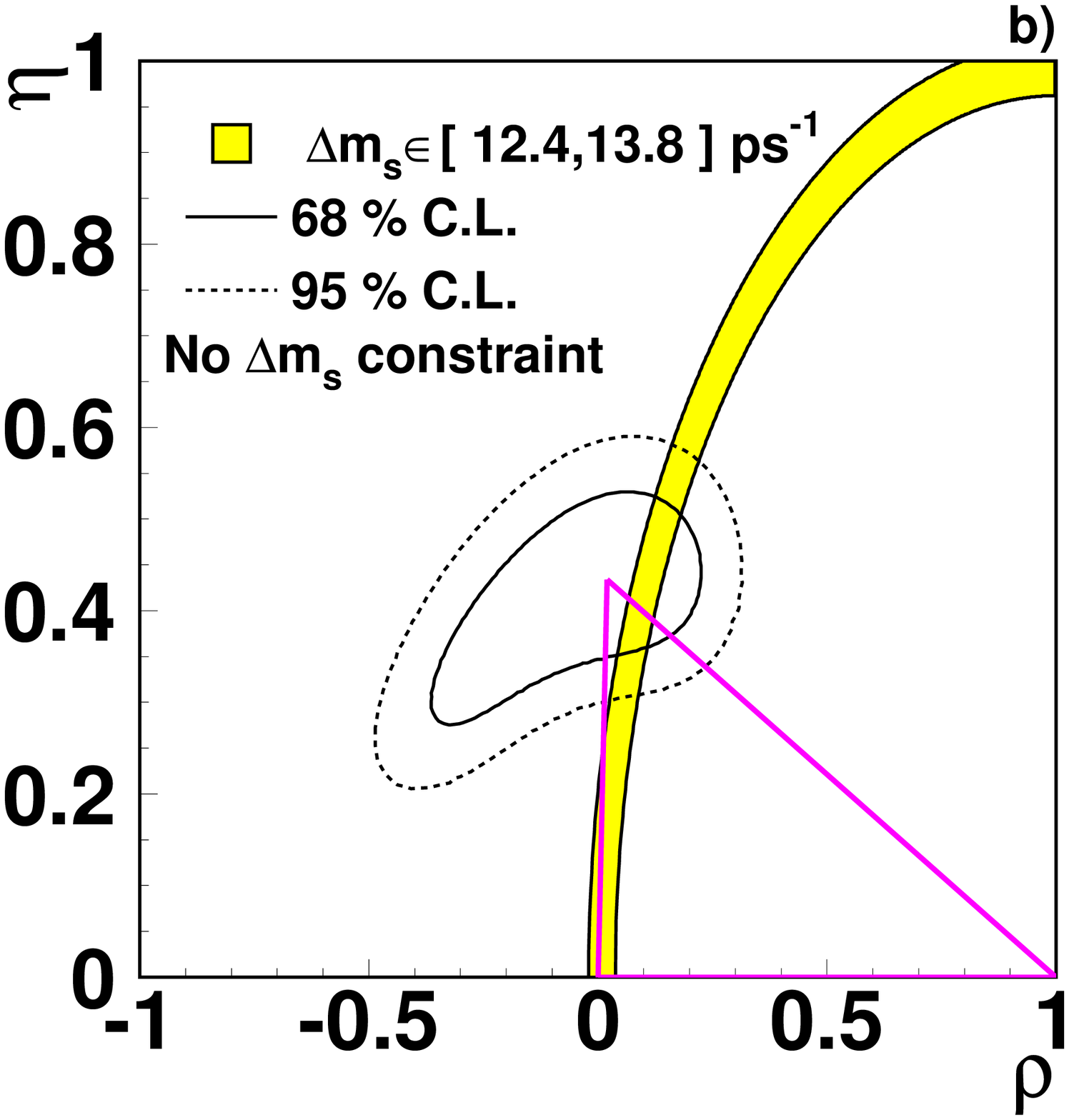}} \\
      \mbox{\includegraphics[width=0.5\figwidth]{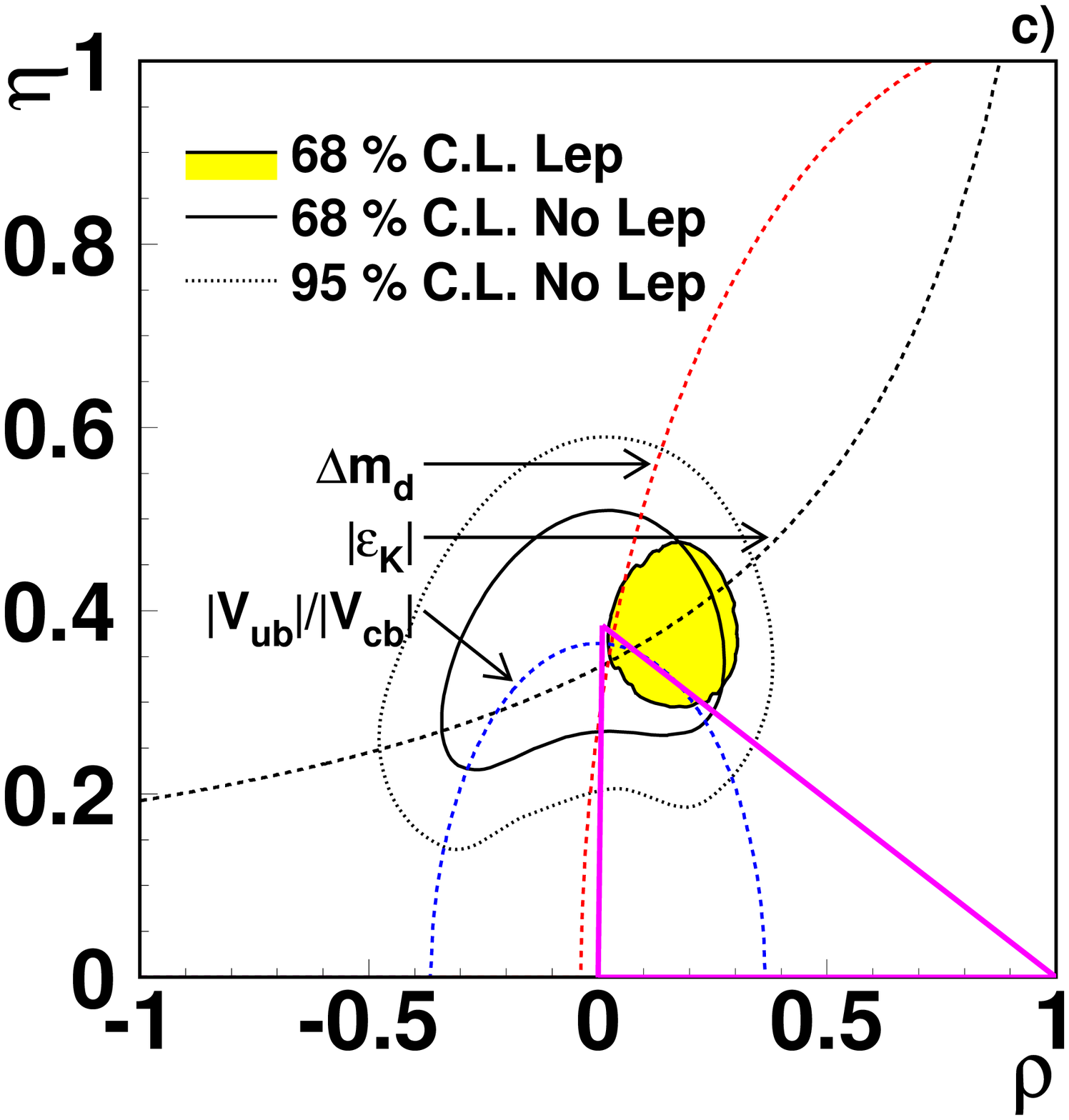}} &
      \mbox{\includegraphics[width=0.5\figwidth]{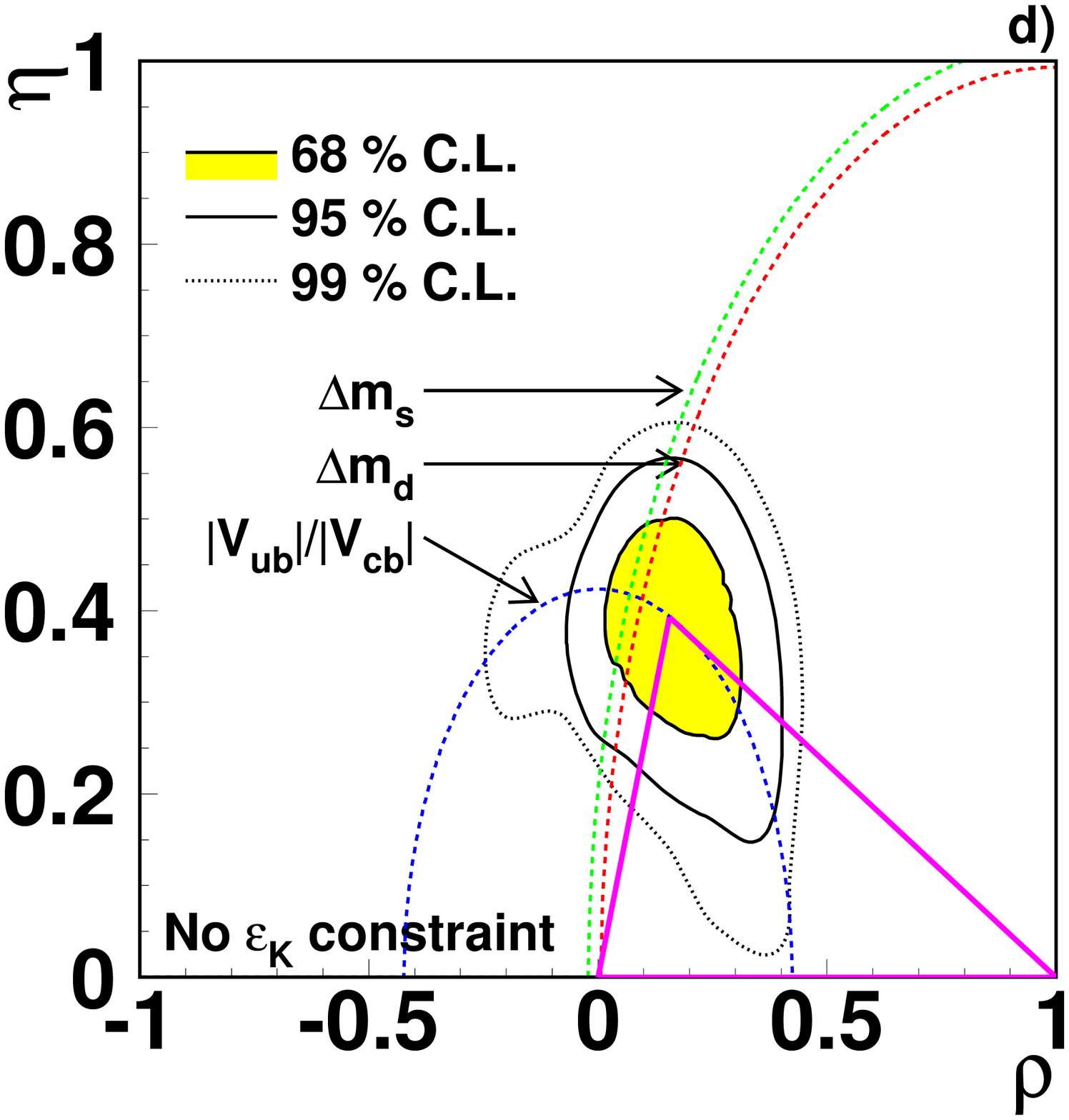}} \\
    \end{tabular}
    \caption{The favoured unitarity triangles and the confidence regions for their vertices
      in the following assumptions: 
      a) the fit using all data described in the text,
      b) the constraint from the $\mathrm{B}^0_\mathrm{s}$ oscillations is not applied,
      c) the LEP measurements are excluded from the fit,
      d) no constraints from the neutral kaon system are applied.
      The band
      in b) displays the values of $\rho$ and $\eta$ corresponding to a value of $\Delta m_s$ 
      between the current lower limit and expected  sensitivity.
      The $\Delta m_s$ limit and the central values of the constraints  are shown 
      in a), c) and d).}
  \end{center}
\end{figure}

\end{document}